\documentclass[12pt,osajnl2,preprint,showpacs,superscriptaddress]{revtex4}  
\usepackage[draft]{hyperref} 
\usepackage{epsfig}
\usepackage{amssymb}
\usepackage{amsmath}

\begin{document}

\title{Self-Chaotization in Coupled Optical Waveguides}

\author{Ramaz Khomeriki}
\author{Archil Ugulava}
\author{Levan Chotorlishvili}
\affiliation{ Department of Exact and Natural Sciences, Tbilisi State University, 3 Chavchavadze, 0128 Tbilisi, Georgia} 

\begin{abstract} 
We consider theoretically two coupled optical waveguides with a
varying barrier height along the waveguides direction. The barrier
could be constructed by the elongated island with a reduced refractive
index (which acts as a potential barrier), such that in the middle region it splits a waveguide into
two weakly coupled parts. It is predicted by numerical simulations
and analytical consideration that the presence of some imperfection of the system parameters can cause splitting of injected laser beam and one will observe two intensity maximums at the output, while for small imperfections the input and output beam intensity distributions will be the
same. The switching between two regimes could be achieved changing spectral width of the beam or refractive index of the island. This nontrivial effect is explained by possibility of transitions between
the different eigenstates of the system in the region of large
potential barrier heights. The mentioned effect could be used for all-optical readdressing and filtering purposes. \end{abstract}

\maketitle

\section{Introduction}
\begin{figure}[ht]
\centerline {\epsfig{file=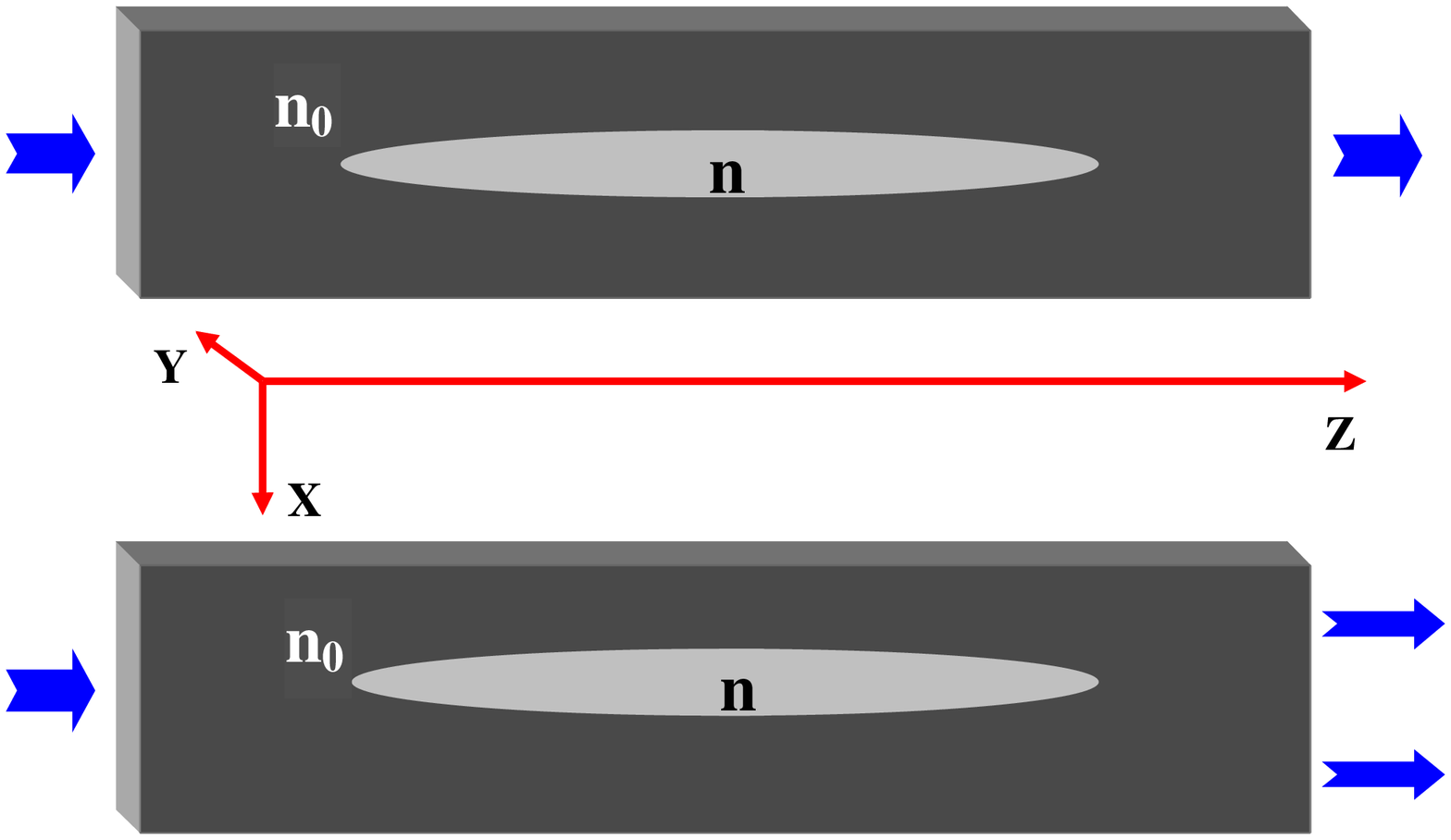,width=8cm}}
\centerline {\epsfig{file=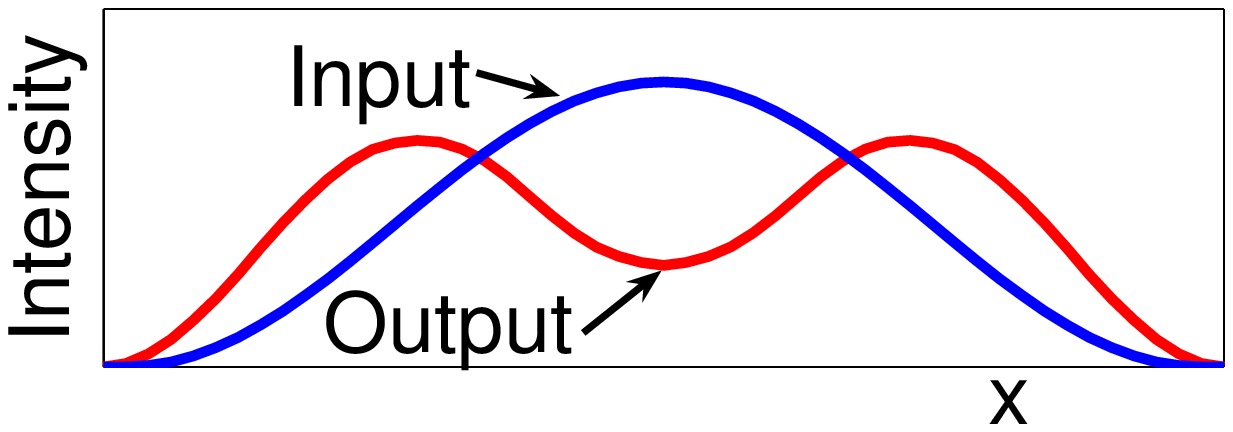,width=8cm}}
\caption{ suggesting experimental setup for observation of self-chaotization process between two eigenstates.
In the proposed device there is an elongated area with a smaller refractive index ($n<n_0$) and spatial length $\Delta z$ splitting the waveguide
into two parts in the middle region. The light is polarized linearly along transverse $y$ direction.
In the upper panel the low imperfections of the system are considered and and as a result the intensity distibutions across the waveguide (axis $x$) is the same both at the input
and output. Lower graph represents the situation for larger imperfections and the picture is different: Instead of a single intensity maximum at the output one should observe splitting of the injected beam. In the lowest graph it is displayed the input-output intensity distributions in case of large imperfections. }
\label{fig:scheme}\end{figure}

The system of coupled optical waveguides \cite{yariv,kivsh-agra}
attracts a steady interest of both theoretical and experimental
researchers since they allow investigating not only purely
optical phenomena (such as anomalous diffraction and reflection
\cite{pertsch}) but also generic problems of nonlinear physics
(creation, propagation and interaction of nonlinear objects
\cite{christ-joseph,eisenberg,mandelik,expe1,yuri,ramaz}, various
types of instabilities \cite{malomed,kivshar1}, nonlinear
bistability \cite{ramaz1}, etc.). At the moment there exists an
increasing interest to observe various quantum effects in coupled
optical waveguides, in other words the people started to think to
use those systems as macroscopic quantum laboratory, as far as the
problem could be described by Schr\"odinger equation, where the waveguide
direction plays a role of time. Here we just quote the prediction
\cite{ramaz2,longhi} and observation \cite{assanto,kivshar2} of
Landau-Zener tunneling, Bloch oscillations
\cite{eisenberg7,elflein}, Anderson Localization \cite{segev} and
analogy of Josephson oscillations \cite{jensen,ramaz3}. However,
all of these studies investigate the dynamical problems, while to
the best of our knowledge no statistical research on the mentioned
systems has been done yet. In the present paper we try to start
filling this gap investigating the influence of the value of the
imperfection of the waveguide and beam parameters on the light propagation through
the waveguide with the longitudinal potential barrier created by
the island with a reduced refractive index (see Fig.
\ref{fig:scheme}). 

Particularly, we propose splitting of the injected laser beam due to imperfections in the system. Indeed, as far as the waveguides are formed by application of external electromagnetic fields (for instance, the elongated island of the reduced refractive index could be formed by a control beam perpendicular to the sample plane as in Ref. \cite{referee}), in this situation due to the photon number fluctuations \cite{RMP} of control beam the refractive index is not homogeneous in space and time. The time averaged stationary picture observed in the experiment could be treated as statistical average over  multiple distributions of random inhomogeneous component of the refractive index.

In the recently proposed scenario of beam splitting \cite{referee} the refractive index defect is placed on the way of the injected signal beam. In other words, moving the defect one can split the beam or only change its direction. In this paper we do not propose to change the position of refractive index barrier, the splitting effect could be achieved by increasing the power of the control beam even in the linear regime of the signal beam. We interpret predicted effects in terms of establishment of statistical equilibrium between symmetric and antisymmetric eigenmodes of the signal beam.

In the figure we present schematics for the observation of the
predicted effect. The system could be effectively considered as a
dynamics in a double well potential with a varying barrier height. We propose to inject the beam at the left described by symmetric (with respect to the center of the waveguide) ground state wavefunction. Then, as it
will be shown below, if the parameter of self-chaotization is
small (i.e. $\delta {\cal K}\Delta z\ll 2\pi$, where $\delta {\cal K}$ is imperfection parameter defined below and $\Delta z$ is a length of the reduced refractive index island, i.e. potential barrier region) one will observe the same intensity distribution at the
right, while for larger self-chaotization parameters the beam will split into
two parts. The same switching effect could be achieved keeping imperfection parameter $\delta {\cal K}$ constant and changing height or length of the potential barrier (i.e. parameters of elongated island with a reduced refractive index) or controlling spectral width of the injected signal beam. 

The explanation of the effect is as follows: initially, when the barrier height is zero (at the left in Fig. 1), the propagation constant of symmetric ground
state wavefunction and the one of first excited antisymmetric
wavefunction differs from one another and no transition occurs,
thus without no barrier, or low barrier heights the input will
coincide with the output in any case. In the region with high
barrier heights the propagation constants of symmetric and antisymmetric
wavefunctions are almost the same, thus even small fluctuations (but in case of large $\Delta z$ for which $\delta {\cal K}\Delta z\geq 2\pi$) can
make transitions between those states and finally thermalize the
situation. In other words the weights of both states will become
the same and at the right one will observe this mixed state (with
two intensity maximums) which will be quite different what we
inject at the left. Obviously the latter scenario does not take
place if imperfections are small, then the transition probability
between the ground and first excited states will be small and the
length of region with nonzero barrier heights is not sufficient
for self chaotization  of the states. Thus one will see the same
picture at the output as at the input.

Thus we have two types of control parameters: one is the value of imperfections and the second is potential barrier characteristics (height and length). Decreasing imperfection parameter one needs more and more barrier height and length in order to observe a splitting effect. And contrary, in presence of large imperfections in the system even small defect with reduced refractive index could split the injected signal beam. It should be especially mentioned that in order not to excite also higher order modes the refractive index barrier variation should be smooth.

\section{Theoretical Model and Numerical Simulations}

Let us start the quantitative consideration with the Maxwell's equations in a non-magnetic medium without free charges, which could be written for the electric field $\vec {\cal E} (\vec r,t)$ as (for the full details of justification of the given procedure and assumed simplifications we redirect reader to the Refs. \cite{christ-joseph,musslimani}): 
\begin{equation}\label{max}
{\vec \nabla}\times {\vec \nabla}\times \vec {\cal E} +
\frac1{c^2}\frac {\partial^2}{\partial t^2}
\left\{ n^2 \vec{\cal E}\right\}=0,
\end{equation}
where the refractive index $n=n(\vec r)$ is allowed to vary in transverse to $y$ plane $xz$ (see Fig. 1 for the axis directions).
Then we may assume a field
polarized and homogeneous along the direction $y$, namely the transverse-electric field reads as
\begin{equation}
\vec {\cal E} (\vec r,t)={\bf\hat e_y}\left({\Psi}(x,z)e^{-i(\omega t-kz)}+c.c.
\right),\end{equation}
where ${\bf\hat e_y}$ is the unit vector of the $y$-coordinate, the stationary envelope function ${\Psi}(x,z)$
depends slowly on the coordinate $z$, $\omega$ is carrier frequency of a beam, $k=\omega n_0/c$ is a
carrier wavenumber and $n_0$ is an average linear refractive index.  In these definitions the Maxwell's equation \eqref{max} becomes
\begin{equation}
2ik\frac {\partial {\Psi}}{\partial z}+\frac {\partial^2{\Psi}}{\partial x^2}  -k^2\left(1-\frac{\left[n(z,x)\right]^2}{n_0^2}\right) {\Psi}=0.
\label{helmoltz}
\end{equation}
\begin{figure}[ht]
\centerline {\epsfig{file=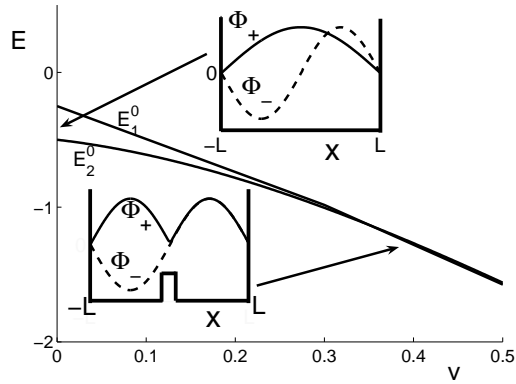,width=7cm}}
\caption{ Dependence of the propagation constants $E_1^0$ and $E_2^0$ of the first two modes on the barrier height $v$ of the double harmonic well potential. As seen, for large $v$-s the eigenvalues $E_1^0$ and $E_2^0$ are very close. Insets display the form of linear modes in double well potential for zero barrier and large barrier heights.  }
\label{mathieu}\end{figure}

Now restricting ourselves to the linear regime, let us assume that $1-n^2/n_0^2$ has a form of square double well potential (considered before in various physical contexts \cite{smerzi,smerzi1,kevrekidis1,kevrekidis2,ober}) plus some random inhomogeneous part ${\cal H}^\prime(z,x)$. In numerical simulations we choose the points in $(x,z)$ plane giving them the random values and then plotting the surface. We run the simulations for the multiple random surfaces averaging afterwards the results.

Defining $\omega_0$ as a central frequency of the injected beam and rescaling to the dimensionless spatial  variables $z\rightarrow zn_0/2\omega_0c$, $x\rightarrow xn_0/\omega_0c$, we arrive to the Schr\"odinger equation written in the form:
\begin{equation}
i\frac{\partial\Psi}{\partial z}=\left({\cal H}_0+{\cal H}^\prime\right)\Psi, \qquad
{\cal H}_0=-\frac{1}{4}\frac{\partial^2}{\partial x^2}+4v(z)V(x)
\label{1}
\end{equation}
where $V(x)$ represents some barrier potential centered at $x=0$; $v(z)$ is a height of the barrier of double square well potential varying along $z$ direction (see for schematics Fig. 1) and dimensionless variable $x$ varies from $-L$ to $L$ (across the whole transversal length of the waveguide), thus $\Psi(z,-L)=\Psi(z,L)=0$. Obviously, in the realistic situation one has a finite well depth resulting in a exponential decay of the wave function at the boundaries instead of the vanishing condition, but this will cause insignificant modification of the results (see e.g. Refs. \cite{smerzi,smerzi1,kevrekidis1,kevrekidis2}) and for simplicity reasoning we will consider further infinite barrier heights at both boundaries of the double well potential. Under these above conditions the eigenvalues of lowest (symmetric with respect to the point $x=0$) and first excited (antisymmetric) eigenstates for various potential heights $v$ are presented in Fig. 2. 

Without random perturbative terms and constant $v(z)$ equation \eqref{1} is easy to analyze and we display dimentionless propagation constants $E_1^0$ and $E_2^0$ for first two linear modes versus barrier height $v$ in Fig. 2.  As seen, if one starts from the symmetric mode and with a zero energy barrier ($v=0$) eventually increasing barrier height, the propagation constants of symmetric and antisymmetric modes become almost equal and infinitesimal fluctuations can cause the transition between those modes. Such a possibility has been discussed earlier \cite{ugulava1,ugulava2} comparing the scenarios of emergence of chaos in classical and quantum pendula. It has been argued that for large barrier heights the energy eigenstates become degenerate due to the presence of fluctuations. Therefore, decreasing the barrier height one will not recover the initial mode, instead one will have a mixed state consisting of both modes with equal weights. In opposite, if one will not increase barrier height enough to consider the states degenerated, the system will recover initial mode.

Our aim is to consider the similar scenario for optical systems like presented in Fig. 1. We will monitor the evaluation of the system for a given random distribution of the potential ${\cal H}^\prime$ in \eqref{1} and then average the evaluation results over many different random distributions, what is presented in Fig. 3. Thus it is assumed that observed stationary picture is a result of averaging over fast time fluctuations of the refractive index distribution in the sample. In case of both a) and b) graphs we take the same initial mode for zero barrier height $\Phi_+=\left(1/\sqrt{L}\right)\cos\left(\pi x/2L\right)$ and the variation of the potential barrier along $z$ is displayed in the inset. In the graph a) the imperfections $\delta{\cal K}$ are small and initial state recovers, while in case of large fluctuations displayed in graph b) one monitors at the output both initial and first excited modes with the equal weights. 

\section{Analytical Consideration}

For analytical treatment we consider first two real symmetric
$\Phi_+$ and antisymmetric $\Phi_-$ eigenfunctions with Eigenvalues $E_1^0$ and $E_2^0$ of the operator ${\cal H}_0$ with constant barrier height $v$. Thus the modes satisfy the following equalities:
\begin{equation}
{\cal H}_0\Phi_+=E_1^0\Phi_+  \qquad {\cal H}_0\Phi_-=E_2^0\Phi_-
\label{2}
\end{equation}
and it is assumed that the following orthonormalization conditions hold:
\begin{equation}
\int\limits_{-L}^L (\Phi_+)^2 dx=1; \quad \int\limits_{-L}^L (\Phi_-)^2 dx=1; \quad \int\limits_{-L}^L \Phi_+\Phi_- dx=0.
\label{3}
\end{equation}
The dependence of Eigenvalues $E_1^0$ and $E_2^0$ of these modes on
the barrier height $v$ is presented in the Fig. 2. In case of zero barrier height the eigenfunctions have trivial form (see the inset in the same figure):
\begin{equation}
\Phi_+=\left(1/\sqrt{L}\right)\cos\left(\pi x/2L\right); \qquad \Phi_-=\left(1/\sqrt{L}\right)\sin(\pi x/L).
\label{4non}
\end{equation}

Presenting now wavefunction $\Psi$ as
\begin{equation}
\Psi(x,z)=\psi_1(z)\Phi_+(x)+\psi_2(z)\Phi_-(x)
\label{5}
\end{equation}
and defining the symmetric ${\cal H}^\prime_s={\cal H}^\prime(z,-x)+{\cal H}^\prime(z,x)$ and antisymmetric ${\cal
H}^\prime_a={\cal H}^\prime(z,-x)-{\cal H}^\prime(z,x)$ parts of the random potential surface ${\cal H}^\prime$ we substitute \eqref{5} into the \eqref{1}. Then multiplying on $\Phi_+$ and $\Phi_-$ and then integrating
over $x$ we get the following set of equations:
\begin{equation}
i\frac{\partial\psi_1}{\partial z}=E_1\psi_1+{\cal K}\psi_2,
\qquad i\frac{\partial\psi_2}{\partial z}=E_2\psi_2+{\cal K}\psi_1
\label{6}
\end{equation}
here $E_{1,2}\equiv E_{1,2}^0+\delta E_{1,2}$ and we have taken into account the following relations and definitions:
\begin{eqnarray}
\int\limits_{-L}^L \Phi_\pm{\cal H}^\prime_s\Phi_\pm dx
=\delta E_{1,2}, \quad \int\limits_{-L}^L \Phi_\pm{\cal
H}^\prime_a\Phi_\pm dx =0, \quad \int\limits_{-L}^L
\Phi_+{\cal H}_0\Phi_- dx =0, \nonumber \\
\int\limits_{-L}^L \Phi_+{\cal H}^\prime_s\Phi_- dx =0,
\quad \int\limits_{-L}^L \Phi_+{\cal H}^\prime_a\Phi_- dx
\equiv{\cal K} \label{7}
\end{eqnarray}
The set of equations \eqref{6} have been considered long before (see e.g. \cite{smerzi,smerzi1}) and has the exact solutions.
Particularly, choosing the initial conditions as $\psi_1(0)=1$ and $\psi_2(0)=0$ and
defining $\Delta E\equiv E_2-E_1$, $\psi_{1,2}=|\psi_{1,2}|\exp\left(i\phi_{1,2}\right)$ we get the solution as follows:
\begin{eqnarray}
|\psi_1|^2=1-|\psi_2|^2, \quad |\psi_2|^2=\frac{4{\cal
K}^2\sin^2\left[(z/2)\sqrt{4{\cal K}^2+\Delta E^2}\right]}{4{\cal
K}^2+\Delta E^2}, \nonumber \\
\phi_2=-\frac{E_1+E_2}{2}z, \quad \phi_1=\phi_2+\phi, \quad
\cos\phi=-\frac{\Delta E|\psi_2|}{2{\cal K}|\psi_1|} \label{8}
\end{eqnarray}
\begin{figure}[ht]
\centerline{\epsfig{file=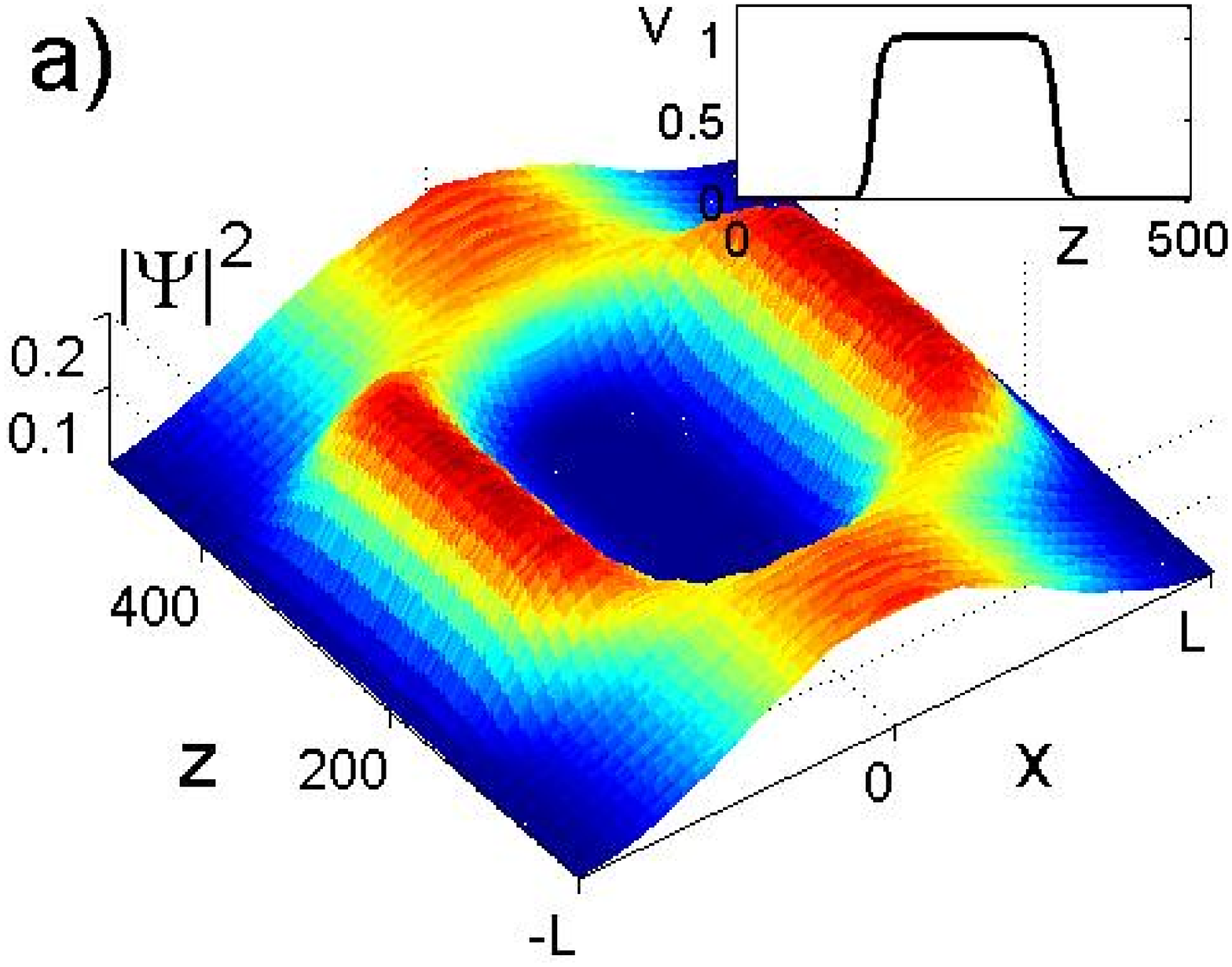,angle=90,width=8cm}\epsfig{file=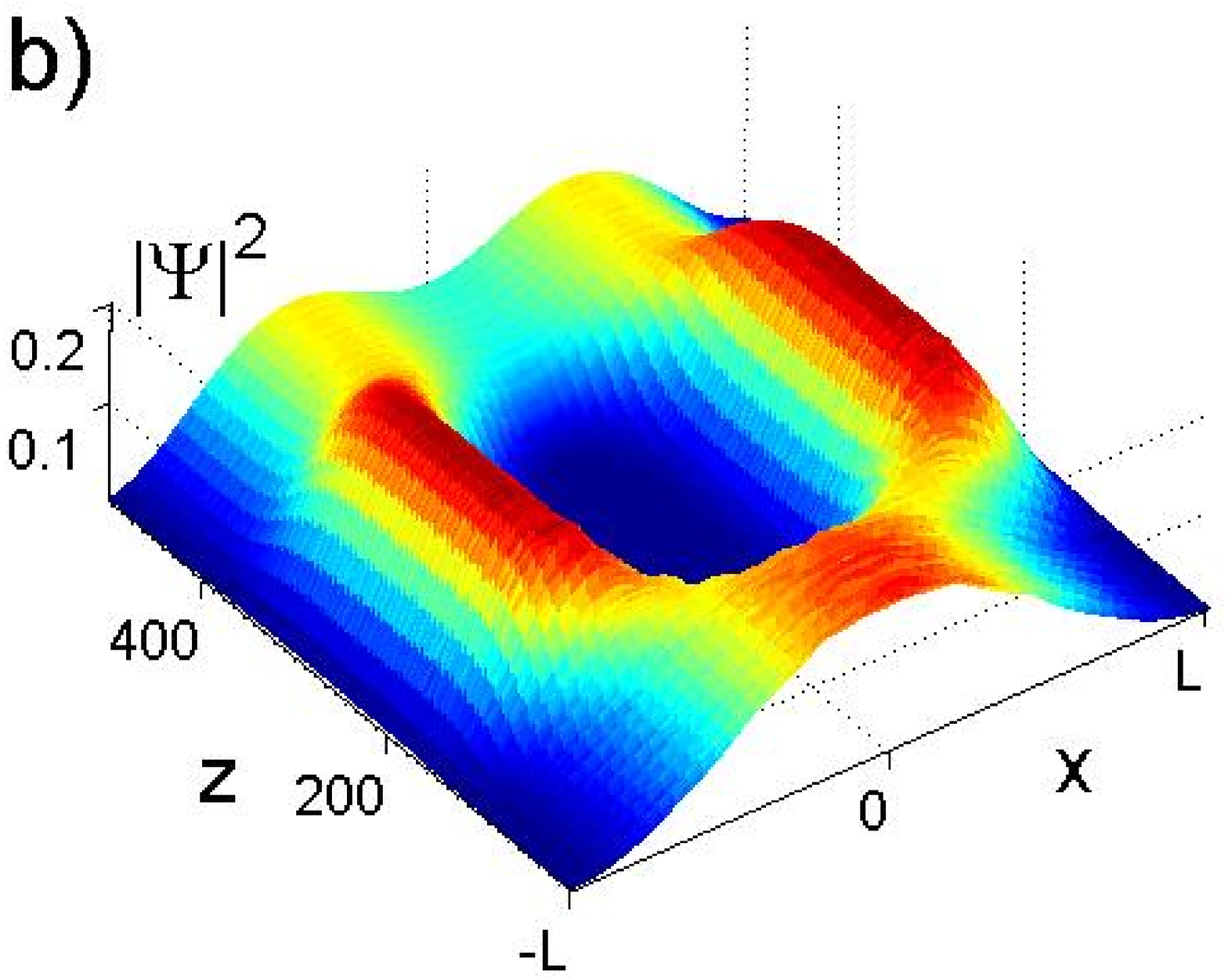, angle=90,width=8cm}}
\caption{ Three dimensional graphs of the intensity distribution describing numerical simulations
on the equation \eqref{1}. The runs are averaged over random distributions
of ${\cal H}^\prime\equiv{\cal H}^\prime_s+{\cal H}^\prime_a$ [see equation \eqref{1}].
Particularly, for the graph a) the distribution is taken
such that $\sqrt{\langle{\cal K}^2\rangle} \approx\sqrt{\langle\delta E_1\rangle}\approx\sqrt{\langle\delta E_2\rangle}\approx 0.0002$, while for the b)
graph $\sqrt{\langle{\cal K}^2\rangle} \approx\sqrt{\langle\delta E_1\rangle}\approx\sqrt{\langle\delta E_2\rangle}\approx 0.002$.
For the larger fluctuations it is clearly seen two intensity maximums at the output part of the graph b). The inset in the graph
a) shows the variation of the potential barrier height $v$
along the waveguide in both large and small fluctuation cases.
 }
\label{fig:dnls}\end{figure}

Now we consider two limiting cases: in the first one corresponding to the small barrier heights $v$, one has strongly nondegenerate situation $\Delta E\gg {\cal K}$, the system remains close to the initial state and $|\psi_1|\simeq 1$ for any $z$. In the second case of large barrier heights $v$ for which $E_1^0-E_2^0\rightarrow 0$ it could be shown that $\delta E_1^0-\delta E_2^0\rightarrow 0$ as well, thus we have the degenerated limit $\Delta E\rightarrow 0$ and from \eqref{8} we get:
\begin{equation}
\psi_1=\cos({\cal K} z) e^{-iE_1 z}, \qquad
\psi_2=-i\sin({\cal K} z) e^{-iE_2 z}
\label{10}
\end{equation}
Thus now the intensity oscillates between the states with a wavenumber ${\cal K}$, which could be treated as a chaotization parameter.

Assuming now that the large barrier heights region lasts for the distance $\Delta z$ and then follows the zero barrier region, from \eqref{5} and \eqref{10} one can write the wave intensity distribution as:
\begin{eqnarray}
&&|\psi|^{2}=|e^{-iE_{1}(z-\Delta z)}\cos({\cal K} \Delta
z)\Phi_{+}-ie^{-iE_{2}(z-\Delta z)} \sin({\cal K}\Delta
z)\Phi_{-}|^{2}= \nonumber \\&&
\cos^{2}({\cal K} \Delta
z)\Phi^{2}_{+}+\sin^{2}({\cal K} \Delta
z)\Phi^{2}_{-}- \sin(2{\cal K}\Delta z)\sin(\Delta E(z-\Delta
z))\Phi_{+}\Phi_{-}.
\label{13}
\end{eqnarray}
As it was expected, expression for intensity contains interferential term.

Let us note that, difference between pure and mixed states is
analogues to the difference appearing in case of interfering coherent
and non-coherent light fluxes, respectively. As well known, in
case of pure states the amplitudes are summed in each point, while in
case of mixed states one has to sum only the intensities. In other words, the interferential term existing in case of pure states will disappear in the mixed states situation by averaging with respect to random phase. 

Let us suppose that transition amplitude ${\cal K}$ is characterized by mean-root-square fluctuation
$\delta{\cal K}\equiv\sqrt{\langle{\cal K}^{2}\rangle}$. We
discuss here two limiting cases corresponding to the small $\delta{\cal
K}\Delta z\ll\pi $ and large $\delta{\cal K}\Delta z\gg\pi$, respectively.
In the first case after averaging with respect to the small
$\delta{\cal K}\Delta z\ll\pi $ we get:
\begin{eqnarray}
\langle\cos({\cal K}z)\rangle\rightarrow 1 , \qquad \langle\cos^{2}({\cal K}z)\rangle\rightarrow 1 , \nonumber \\ 
\langle\sin({\cal K}z)\rangle\rightarrow 0 , \qquad \langle\sin^{2}({\cal K}z)\rangle\rightarrow 0 , \nonumber \\ 
\langle\Psi\rangle\rightarrow\Phi_{+}(x), \qquad
\langle|\Psi|^{2}\rangle\rightarrow\Phi^{2}_{+}(x). 
\end{eqnarray}
It is easy
to see that initial wave function in the "input" is completely
revived in the "output". In case of large quantity $\delta{\cal
K}\Delta z\gg\pi$ situation is completely different. In this case
error incursion $\delta{\cal K}$ up to
the values $2\pi$ takes place on the distance $\Delta z $ and this leads to the lost of information about the systems initial state. After averaging
procedure for wave function we have $\langle\Psi\rangle=0 $ and
taking into account expressions 
\begin{equation}
\langle\cos^{2}({\cal K}\Delta z)\rangle=\langle\sin^{2}({\cal
K}\Delta z)\rangle=\frac{1}{2}, \quad
\langle\sin(2{\cal K}\Delta z)\rangle=0,
\end{equation}
we obtain for intensity the following:
\begin{equation}
\langle|\Psi|^{2}\rangle=\frac{1}{2}\left(\Phi^{2}_{+}(x)+\Phi^{2}_{-}(x)
\right). 
\end{equation}

So, in this case system's state is mixed state in which symmetric
and anti-symmetric components are presented with equal
probabilities. It is important that, condition of
self-chaotization $\delta{\cal K}\Delta z\geqslant\pi$ could hold
even for small $\delta{\cal K}$ increasing the distance $\Delta z$.

\section{Conclusions}

Summarizing we have investigated the influence of the system imperfections on the intensity distribution along the waveguide direction.
The predicted splitting of the beam for large imperfections is explained in
terms of the chaotized transitions between the quasi-degenerated eigenmodes of the Schr\"odinger equation in double well potential. Thus one can control the splitting process by multiple ways: either changing the imperfection value of the system or the spectral width of the injected beam or vary the height and/or length of the barrier potential. In presence of large imperfections in the system even small defect with reduced refractive index could split the injected signal beam.

\paragraph{Acknowledgements.} The designated project has been fulfilled by financial support of the Georgian National Science Foundation (Grant No GNSF/STO7/4-197). R. Kh. acknowledges support by Marie-Curie Foundation (contract No MIF2-CT-2006-021328) and USA CRDF (award No GEP2-2848-TB-06).

\end{document}